%% file: bubrevised3.tex
\documentclass{JHEP3}
\usepackage{epsfig}
\usepackage{latexsym}
\usepackage{epsfig,amssymb,euscript}
\usepackage{amsmath}
\input texpref.tex

\newcommand{\be}{\begin{equation}}
\newcommand{\ee}{\end{equation}}
\newcommand{\bea}{\begin{eqnarray}}
\newcommand{\eea}{\end{eqnarray}}
\newcommand{\bean}{\begin{eqnarray*}}
\newcommand{\eean}{\end{eqnarray*}}
\newcommand{\nn}{\nonumber}
\newcommand{\ft}[2]{{\textstyle\frac{#1}{#2}}}

\newcommand\D{\mathcal{D}}

\newcommand{\de}{\partial}

\preprint{CERN-PH-TH/2004-252}

\title{Bubbling AdS$_3$}

\author{D. Martelli and J.F. Morales\\

\vspace{0.3 cm}

\parbox[t]{6in}
{Department of Physics, CERN Theory Division,
1211 Geneva 23, Switzerland.}\\

\vspace{0.5 cm}

\email{Dario.Martelli, Francisco.Morales.Morera@cern.ch}
}

\abstract{In the light of the recent Lin, Lunin, Maldacena (LLM)
results, we investigate $\ft12$-BPS geometries in minimal (and
next to minimal) supergravity in $D=6$ dimensions.
In the case of
minimal supergravity, solutions are given by
fibrations of a two-torus $T^2$
specified by two harmonic functions.
For a rectangular torus the two functions are related by a
non-linear equation with rare solutions: $AdS_3\times S^3$, the
pp-wave and the multi-center string. ``Bubbling'', i.e.
superpositions of droplets, is accommodated by allowing the complex
structure of the $T^2$ to vary over the base.
The analysis is repeated in the presence of a tensor multiplet and
similar conclusions are reached, with generic solutions describing
D1D5 (or their dual fundamental string-momentum) systems.
In this framework, the profile of the dual
fundamental string-momentum system is identified with the boundaries
of the droplets in a two-dimensional plane.}

\begin{document}

\section{Introduction}

The AdS/CFT correspondence \cite{malda} relates deformations of
AdS geometries to states  in the boundary CFT.
 A concrete realization of this idea was recently proposed by
 Lin, Lunin, Maldacena (LLM) in \cite{LLM}, where
classical geometries of Type IIB supergravity corresponding to
$\ft12$-BPS states in ${\cal N}=4$ SYM were constructed. On the
gauge theory side the solutions correspond to chiral primary
operators with conformal weight $\Delta=J$ and are dual to
deformations of $AdS_5\times S^5$ (or the pp-wave) backgrounds
preserving half supersymmetry. The field theory states were
previously found to have a description in terms of free fermions
\cite{bere,corley}. In a semiclassical limit these states can be
depicted as ``droplets'', or ``bubbles'', on a two-dimensional
plane, which is the phase space of these fermions. The remarkable
result of \cite{LLM} was to show that to droplet configurations 
correspond supersymmetric solutions of Type IIB supergravity, with
$SO(4)\times SO(4)$ isometry. The geometries are completely
specified by a distribution of charge $z=\pm \ft12$ on a two
dimensional plane non-trivially embedded in space-time. In the
geometry the ``bubbles'' define islands in space where one of the
two $S^3$ shrinks to zero size and only for $z=\pm \ft12$ the
corresponding geometries are regular.

The results in \cite{LLM} provide the most general supersymmetric
solutions of Type IIB supergravity, in the presence of a five-form
flux admitting an $SO(4)\times SO(4)$ group of isometries\footnote{In \cite{LLM}
a similar  analysis was applied to M-Theory geometries, and constituted an
extension of the results of \cite{paper1}.}.
This analysis was done using the techniques first introduced in
\cite{GMPW} (and subsequently exploited in \cite{GGHPR}
%,GP,GG,
-\cite{GMR}
and many others) and is greatly simplified by the large amount of
isometry. The field theory duals are given by SYM states
satisfying
 $\Delta=J$ built as multi-trace products
 of $J$ scalar fields of a single specie.

The scope of this note is to investigate, from the supergravity
point of view, the similar story in $D=6$. Solutions will
correspond to $\ft12$-BPS deformations of $AdS_3\times S^3$ (or
its pp-wave limit) and are dual to chiral primaries
in the boundary CFT. The existence of a 
 free fermion description of primaries in the two-dimensional
CFT \cite{Lunin:2002fw} suggests that bubbling solutions
should find  room in six-dimensional supergravity.
 Here we show that this is indeed the case.
Much is already known about the supergravity 
description of chiral
primaries of the two-dimensional CFT \cite{paradox,LMS,LMM}, and
we will ask ourselves whether these results can be reinterpreted
in terms of bubblings of $AdS_3$.

Half BPS geometries associated to excitations around $AdS_3\times
S^3$ are dual to chiral primaries in the two-dimensional D1D5 (or
FNS) CFT. The spectrum of chiral primaries and its dual KK
descendants in supergravity have been worked out in
\cite{deboer1}
%,deboer2,ghmn1,
-\cite{ghmn2}. States in the CFT are classified by four
charges $h,\bar{h},j,\bar{j}$ describing the quantum numbers under
the isometry group $SO(2,2)\times SO(4)\sim SL(2,R)_L\times
SL(2,R)_R\times SU(2)_L\times SU(2)_R$. $h,\bar{h}$ describe the
conformal dimension of the two-dimensional  CFT and $j,\bar{j}$
the R-symmetry charges. In the case of minimal ${\cal N}=(1,0)$
supergravity in $D=6$, the CFT has
 ${\cal N}=(4,0)$ supersymmetry. This sector is universal to any
 supergravity in $D=6$ and solutions are shared by
 supergravities following from reductions on $T^4$, $K3$ and
 orientifolds.

 In analogy with the ten-dimensional case we start by decomposing
the isometry group as $SO(2)^2\times SO(2)^2_{\theta_1,\theta_2}$
and consider states with zero $SO(2)^2_{\theta_i}$
charges i.e. $h=\bar{h}$ and $j=\bar{j}$. $\ft12$-BPS states
correspond to chiral primaries $h=j$ and therefore we look for
states with $h=\bar{h}=j=\bar{j}={m\over 2}$. There is a single
state of this type for each $m$ in the spectrum of KK descendants
of the gravity multiplet and one for each tensor
multiplet\footnote{This can be easily seen from the list (3.1) in
\cite{ghmn1,ghmn2} for KK descendants of the various ${\cal N}=(1,0)$
supermultiplets.}.
 We therefore look for solutions in the pure ${\cal N}=(1,0)$
 supergravity and its minimal extension by
 adding a tensor multiplet.

Notice that, contrary to the ten-dimensional case studied in
\cite{LLM}, requiring the solution to admit an $SO(2)\times SO(2)$
group of isometries \emph{does not} fix uniquely the form of the
internal space. In particular, on the two-torus $T^2$ we could have
a non-diagonal metric, and generically a non-trivial fibration
structure. We start our analysis being conservative, working in
the minimal supergravity, and mimicking the ansatz used in
\cite{LLM} with $T^2=S^1\times S^1$. Rather surprisingly, we find
an almost identical set of equations describing our solutions. In
particular, it turns out that a function $z$ obeys the same
equation as in \cite{LLM}. However, unlike in the LLM case, the
 Bianchi identity translates into a further harmonic condition on the function
 $h^2$, related via a non-linear equation to $z$.
 The important property of
linearity of the equations is in this way lost and solutions are
rare: $AdS_3\times S^3$, the pp-wave and the multi-center string.

It turns out that the resolution of this problem arises from
relaxing the initial metric ansatz, namely, considering a torus
which is not any more rectangular.  Indeed, using the more general
form of $\ft12$-BPS solutions of minimal  supergravity given in
\cite{GMR}, we show then how bubbling can be accommodated by
 allowing for a tilted $T^2$. In this case, the non-linear
 relation between $z$ and $h^2$
 is lifted, and one is able to freely
superpose different solutions in a fashion similar to \cite{LLM}.
The resulting geometries are given in terms of harmonics generated
by lines of charges distributed along the boundary of droplets in
a two-dimensional plane. The cycles of the torus degenerate
along this plane, while crossing the charged strings the
corresponding pinching cycles get flipped.

Finally we extend our analysis by adding a tensor multiplet to the
minimal theory, namely an anti--self--dual three-form, and a
scalar field. This theory includes a wider class of D1D5 classical
geometries like for instance giant gravitons, and have been
systematically studied in \cite{LMS}. The familiar string profiles
describing these solutions are reinterpreted here as the
boundaries of the droplet configurations in the two-dimensional
plane.

The paper is organized as follows: In section \ref{12bps} we
describe the solutions for minimal supergravity in $D=6$. We start
by considering a simple metric ansatz where the torus of isometries
is rectangular. In section \ref{GMRform} the solutions are written in
the canonical form of \cite{GMR} and the ansatz for the metric is
relaxed to accommodate bubblings. In section \ref{minbub} we
discuss general features of bubbling solutions.
In section \ref{addtensor} we add a tensor multiplet and
discuss bubblings in this extended framework. 
Finally, in section \ref{conclusions} we draw some conclusions.

\vskip 1cm
\noindent {\bf Note added:} While this work was being completed, the
paper \cite{LVW} appeared, which overlaps with the results in 
our section \ref{resume}.

\section{$\ft12$-BPS solutions in minimal supergravity}
\label{12bps}

\subsection{The supersymmetry conditions}

\label{minimal}

In this section we make use of the results of \cite{GMR} to find
$\ft12$-supersymmetric solutions of minimal ${\cal N}=(1,0)$ supergravity
in 6 dimensions of the type recently constructed in \cite{LLM}
\footnote{The results in this section were derived in collaboration with
G. Dall'Agata.}.
We use the six-dimensional conventions of \cite{GMR}, and adhere
to the notation of \cite{LLM}.

Minimal supergravity in 6 dimensions comprises a graviton $g_{mn}$,
a two-form $B^+_{mn}$ with self-dual field strength, and a symplectic
Majorana--Weyl gravitino $\psi_\mu^A$.
The Killing spinor equation reads:
\begin{equation}
 \nabla_m \epsilon - \frac{1}{4}G_{mnp}\gamma^{np} \, \epsilon=0
\label{killing}
\end{equation}
where $G=dB^+$ is self--dual and $\epsilon$ is
symplectic-Majorana--Weyl, i.e. it has 8 real degrees of freedom.

All supersymmetric solutions of minimal supergravity were characterized
in \cite{GMR} in terms of a vector $V$ and a triplet of self-dual
three--forms $X^i$ constructed in terms of spinor bi-linears.
These objects satisfy some algebraic constraints, following from Fierz
identities, and some differential conditions which are equivalent to
the supersymmetry equation (\ref{killing}).
For instance, one of the algebraic constraints implies that the vector
$V$ is null
\begin{equation}
V_{m}V^{m} = 0~.
\label{eq:null}
\end{equation}
The Killing spinor equation (\ref{killing}) is then equivalent to
the following differential conditions
\begin{eqnarray}
\nabla_m V_n & = & V^p G_{pmn}\label{vsusy}\\
\nabla_m X^i_{npq} & = & G_{mp}{}^{r} X^i_{rqn}  + G_{mq}{}^{r} X^i_{rnp} +
G_{mq}{}^{r} X^i_{rpn}~.\label{xsusy}
\end{eqnarray}
Notice that (\ref{vsusy}) implies that $V$ is a Killing vector.
We refer to \cite{GMR} for further details concerning the algebraic and
differential conditions.

In analogy with \cite{LLM},  we now introduce ansatze for the metric and the
self-dual 3-form that admit a $U(1)_{\theta_1} \times U(1)_{\theta_2}$
group of isometries
\bea
ds^2 &=& g_{\mu\nu}\, dx^\mu dx^\nu -e^{H+G} \, d\theta_1^2-
e^{H-G} \, d\theta_2^2\nn\\
G &=&F \wedge d\theta_1+\tilde{F} \wedge d\theta_2
\label{fluxansatz}
\eea
with $\mu=0,1,2,3$.
Self-duality of the three-form\footnote{Hopefully it should be clear when $G$
denotes the three-form and when it denotes a function.}
 field strength $G$ yields the following relations:
\begin{equation}
 F\equiv dB=e^{G}\, *_4\, \tilde{F} \qquad   \tilde{F}\equiv
dB= - e^{-G}\,   *_4\, F
\end{equation}
where $*_4$ is the Hodge star operator with respect to the metric
$g_{\mu\nu}$. We would like now to use the constraints imposed by
supersymmetry in order to determine the functions entering in the
metric  as well as the two-form $F$ in (\ref{fluxansatz}). 
This can be easily achieved using the results of \cite{GMR}.
Although in this latter reference it was introduced a
null orthonormal frame adapted to the Killing vector $V$,
we find it more convenient to adapt the
conditions in \cite{GMR} to our metric ansatz. In particular, we
will first extract from the six-dimensional bi--linears a set of
four--dimensional forms.

The null Killing vector $V^{m}$ can be always chosen of the form $V^{m} =
(K^{\mu},f_1,f_2)$.
Since $\nabla_{(m} V_{n)}= 0$, we also have that $K^{\mu}$ is a Killing vector,
as well as that $\partial_{\mu}f_1=\partial_{\mu} f_2 = 0$.
We can therefore normalize our vector so that
\begin{equation}
V^m = (K^\mu,1,1)~, \qquad V_m = (K_\mu, - e^{H+G}, - e^{H-G})~.
\label{eq:nulldeco}
\end{equation}
As $V$ is null, it follows that the vector $K$ is timelike, and its norm is given by
\begin{eqnarray}
K \cdot K & = & e^{H+G} + e^{H-G}\equiv h^{-2}~, \label{KK}
\end{eqnarray}
hence we can choose a time coordinate $t$ such that the metric takes the
form
\begin{equation}
ds^2 = h^{-2}(dt+C)^2 -g^3_{ij}dx^i dx^j  - e^{H+G} \,d\theta_1^2 - e^{H-G} \,d\theta_2^2
\end{equation}
whence
\begin{eqnarray}
K & = & \partial / \partial t \quad{\mathrm{as~a~vector}}\\
K & = & h^{-2} (dt+C)\quad{\mathrm{as~a~one-form}}
\end{eqnarray}
$C=C_i \, dx^i$, $i=1,2,3$ and of course nothing depends on
$t,\theta_1$ or $\theta_2$.
To proceed, we define a set of forms by decomposing the 3--forms
$X^{i}$
\begin{equation}
X^{i}_{\mu\theta_1\theta_2} = L^i_{\mu}\,, \quad X^{i}_{\mu\nu \theta_1} =Y^{i}_{\mu\nu}\,,
\quad X^{i}_{\mu\nu\theta_2} = \tilde Y^{i}_{\mu\nu}\,, \quad
X^{i}_{\mu\nu\rho}= \tilde L^i_{\mu\nu\rho} .
\label{eq:Xdeco}
\end{equation}
Due to the self-duality of $X^i$, these are not all independent (see
appendix \ref{appe} for details).
It turns out that the differential equations that these forms satisfy
determine the complete form of the metric and self-dual three-form $G$.
We have relegated the detailed derivation in the appendix \ref{appe}.

\subsection{General solution for a rectangular torus}
\label{resume}

The final result is the following.
The metric is specified by a single function $G$ and is given by:
\begin{eqnarray}
ds^2 &=& h^{-2} (dt+C)^2-h^2 (dy^2+dx_1^2 + dx_2^2)-
y e^{G} \,d\theta_1^2 - y e^{-G} \,d\theta_2^2 \label{themetric}\\
h^{-2} &=& 2y\cosh G\label{defh2}\\
dC & =& \frac{1}{y} *_3 dz \label{dCequation} \qquad \quad z  =
\frac{1}{2} \tanh G\label{defz}
\end{eqnarray}
where $*_3$ is the Hodge star in the flat metric
$ds^2_3=dy^2+dx_1^2+dx_2^2$.
 Notice that $z$ and $h^2$ are defined \emph{exactly} as in \cite{LLM}.
The three-form is given by
\begin{eqnarray}
%G &=& F \wedge d\theta_1 +\tilde{F}\wedge d\theta_2\nonumber\\
F&=& dB_t\wedge (dt+C)+B_t dC+d\hat{B}\nonumber\\
\tilde F &=& d\tilde{B}_t\wedge (dt+C)+\tilde{B}_t dC+d\hat{\tilde{B}}\nonumber\\
B_t &=&\frac12\, y e^{G}   \qquad \tilde{B}_t=\frac12\, y e^{-G}\nonumber\\
d\hat{B} &=& -d\hat{\tilde{B}}=\frac{1}{2} \, y\, *_3 d h^2\label{theflux}~.
\end{eqnarray}
Consistency of equations (\ref{defz}) and (\ref{theflux}) requires
that
 $d(dC)=0$ and $d(d\hat{B})=0$. These impose
{\em two} second order equations to be satisfied by $h^2$ and $z$:
\begin{eqnarray}
&& \Delta_3 z -\frac{1}{y}\partial_y z  =  0\label{z6d}\\
&& \Delta_3 h^2 +\frac{1}{y}\partial_y h^2  =  0
\label{d2}
\end{eqnarray}
with $\Delta_3 = \partial_y^2+ \partial_1^2+\partial_2^2$.
 Recalling that $h$ and $z$ are related by the equation
\begin{eqnarray}
h^2   & = & \frac{1}{y} \sqrt{\frac{1}{4}-z^2}
\label{hvsz}
\end{eqnarray}
we see that eqs. (\ref{z6d}) and (\ref{d2}) are two differential
equations on a single function $z$. This is substantially
different from \cite{LLM} where the equations $d\hat{B}=0$ and
$d\hat{\tilde{B}}=0$ were automatically satisfied for a function $z$
obeying (\ref{z6d}).

Notice that like in \cite{LLM}, solutions are
regular for $z=\pm \ft12$ on  the two dimensional plane $y=0$.
Indeed, only for these values of $z$, the shrinking $S^1$ at $y=0$
combine with $y$ to reconstruct a regular (i.e. free of conical singularities)
$\R^2\times S^1$.  Therefore, exactly like for LLM, non-singular solutions
are completely specified by two-dimensional figures (on the $y=0$ plane)
representing regions where $z=\pm{1\over 2}$.
However, solutions to (\ref{z6d}-\ref{hvsz}) are now
sporadic. Remarkably, we find
that $AdS_3\times S^3$, the pp-wave  and the multi-center string
do satisfy these equations. They correspond to the simplest
figures: the disk, the upper half-plane and points (or small
far away droplets). We will refer to the filled figures as
droplets.
The functions $z$ and $h^2$
specifying these solutions were given in \cite{LLM} and will be
displayed below momentarily.

Equations (\ref{z6d}) and (\ref{d2}) can be expressed as
Laplacian equations in $d=6$ and $d=4$
\begin{eqnarray}
 \Delta_6 \, \left(\frac{z}{y^2}\right) &= &0\label{harm6}\\
\Delta_4 \, h^2 & = &0\label{harm4}
\end{eqnarray}
where $y$ is interpreted as the radius of the extra $S^3$ and
$S^1$ auxiliary spheres respectively.
The function $z$ and the one-form $C$
can be written in terms of integrals over the boundary of
the corresponding droplets in the $y=0$ plane \cite{LLM}:
\bea
z(x_1,x_2,y) &=&
{y^2\over \pi} \, \int_\D {z(x_1',x_2',0)\, dx'_1 dx'_2\over
( |{\bf x}-{\bf x'}|^2+y^2)^2}=
\sigma-{1\over 2 \pi}\, \int_{\partial \D} dv |\partial_v {\bf x'}|\,
 { {\bf n}\cdot ({\bf x}-{\bf x'}(v)) \over
|{\bf x}-{\bf x'}(v)|^2+y^2} \label{zc}\\
 C_i(x_1,x_2,y) &=&
{\epsilon_{ij}\over \pi} \, \int_\D {z(x_1',x_2',0)\,(x_i-x'_i)\,
 dx'_1 dx'_2\over
( |{\bf x}-{\bf x'}|^2+y^2)^2}=
{\epsilon_{ij}\over 2 \pi}\, \int_{\partial \D} dv
 { \partial_v x'_j(v) \over
|{\bf x}-{\bf x'}(v)|^2+y^2} \label{zc2}
\eea
In the right hand side of the equations we have introduced a
parametrization of the one-dimensional boundary $\partial \D$ of
the droplets in the $y'=0$ plane.
 From (\ref{zc2}) we see that
$C_i$ are harmonic functions in $d=4$ generated by a one-dimensional
charge distribution along the line $(x'_1(v),x'_2(v))$  parametrized by $v$,
with charge density $\partial_v x'_j(v)$. $\sigma =\pm {1\over 2} $ is a
contribution coming from infinity arising for solutions for which
$z=\pm {1\over 2}$
outside some circle of large radius \cite{LLM}.

Let us now consider the function $h^2$. Since this is
a harmonic function in $d=4$, the general solution is specified by
some source distribution $\rho$, and can be written as
%is generated by the
%same charge distribution (with different density) in the four-dimensional space.
%The reason for this, will become clear later.
 % when we embed this four-dimensional
%space in our spacetime by identifying the auxiliary angle with a linear
%combination of the $\theta_i$'s.
%Invariance of our solution
%under $\partial_{\theta_i}$ requires then that charges sit at $y=0$.
%Then we can write
\be
h^2(x_1,x_2,y) = \int_\D {\rho(x_1',x_2')\,
 dx'_1 dx'_2\over
 |{\bf x}-{\bf x'}|^2+y^2} ~.\label{h2de}
\ee
Note that this expression takes into account the fact that $h^2$ must be
invariant along the auxiliary $S^1$, so that  the density
$\rho(x_1',x_2')$ should sit on the $y'=0$ plane. The physical reason for this
will become clearer later.
 %describing the "mass" density.
Interestingly, this density can be computed explicitly
 in the three cases at hand where $h^2$ is related to $z$ via the non-linear
 relation (\ref{hvsz}). It turns out that in all these cases
 we can write $h^2$ as the following boundary integral
 %\footnote{Here we always
 %focus on near horizon geometries. For asymptotically flat solutions one
 %replaces $h^2\to h^2+1$.}:
\be
h^2(x_1,x_2,y) = \frac{1}{2\pi}\int_{\partial \D} {dv\over
 |{\bf x}-{\bf x'}(v)|^2+y^2}~.\label{h2v}
\ee
This  will be shown explicitly below for the basic figures: the circle,
half-plane and points. One can wonder whether formulas (\ref{zc},\ref{h2v}) can be extended to more
complicated figures. The problem is that, if one tries to do so, the resulting
solutions will fail to obey (\ref{hvsz}).
In the next section we will show how this can be solved by relaxing
the metric ansatz.

Finally, it is interesting to notice that it is also possible to trade
the linear equation for $h^2$ in favor of a non-linear one for
$z$, which reads
\begin{equation}
(\nabla z)^2  =  \frac{4}{y^2} (\frac{1}{4}-z^2)^2\label{iconal}~.
\end{equation}

\subsubsection*{Examples:}

Here we collect the form of $z$ and $h^2$ for the simplest
solutions (the only known to us) to the system (\ref{z6d}-\ref{hvsz}).
%We have defined $\vec{x}=(x_1,x_2)$.

\begin{figure}[t]
\epsfxsize = 14cm
\centerline{\epsfbox{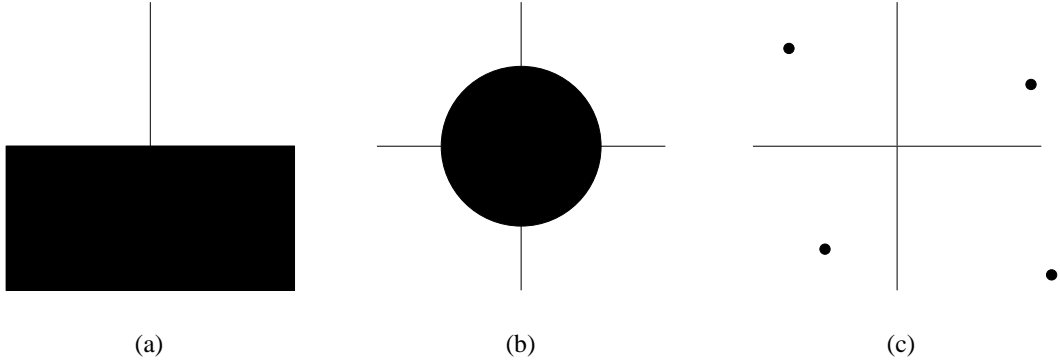}}
\caption{The basic figures in minimal supergravity. (a) The pp-wave. 
(b) $AdS_3\times S^3$. (c) Multi-center string.}
\label{3figs}
\end{figure}

\vskip 4mm
\noindent
{\bf pp-wave}

\begin{eqnarray}
z &=&\frac{1}{2}{x_2\over \sqrt{x_2^2+y^2}} \nonumber\\
h^2 &=&  \frac{1}{2}{1\over \sqrt{x_2^2+y^2}}~.
\end{eqnarray}
This corresponds to dividing the $y=0$ plane in two regions (filled and empty),
separated by the $x_1$ axis \cite{LLM}.
 The functions $z$ and $h^2$ can be written as the integrals (\ref{zc},\ref{h2v})
 over the $x_1$-axis dividing the two regions:
\be
 {\bf x}'(v)=(v, 0)\quad\qquad -\infty< v<\infty~.
\ee

\noindent
{\bf $\mathbf{AdS_3\times S^3}$}

\begin{eqnarray}
z &=&\frac{1}{2}{{\bf x}^2+y^2-a^2 \over \sqrt{({\bf x}^2+y^2+a^2)^2-4 \,a^2\,
{\bf x}^2}}
 \nonumber\\
h^2 &=&  { a \over \sqrt{({\bf x}^2+y^2)^2-2 a^2
({\bf x}^2-y^2)+a^4}}~.
\label{round}
\end{eqnarray}
This corresponds to a round disk of radius $a$ centered in the origin \cite{LLM}.
The  functions $z$ and $h^2$ can be written as the integrals over
the droplet boundary (a circle of radius $a$) parametrized by $v$:
\be
 {\bf x}'(v)=(a \cos \, v, a \sin \, v)\quad\quad 0< v<2 \,\pi~.
\ee

\noindent
{\bf Multi-center string}

\begin{eqnarray}
h^2 &=& {1\over \lambda^2}\, H \quad\quad
H=\sum_i {Q_i\over (\vec{x}-\vec{x}_{0,i})^2+y^2}\nonumber\\
z &=&  \pm \sqrt{ \frac{1}{4}-{1\over \lambda^4} \, H^2 y^2}\quad \quad
\lambda\to \infty\label{eqsmulti}~.
\end{eqnarray}

\noindent
In this case the equation for $z$ is satisfied in the
limit $\lambda\to \infty$. Suppose we take the plus sign in
(\ref{eqsmulti}),
then in the limit, $e^{-G}\sim yH/\lambda^2$. Substituting these into
the metric one finds $C\approx 0$ and
\begin{equation}
ds^2= H^{-1}(d\tilde{t}^2-dw^2) -H
(d\vec{\tilde{x}}^2+ d\tilde{y}^2+\tilde{y}^2 d\theta_2^2)
\end{equation}
 where we have rescaled $\tilde{t}=\lambda\,  t, w=
 \lambda\, \theta_1,
\tilde{y}=\lambda^{-1}\, y$, $\vec{\tilde{x}}=\lambda^{-1}\,\vec{x}$.
The resulting solution corresponds to a multi-center string in $D=6$.
Obviously, the same solution is obtained choosing the minus sign in (\ref{eqsmulti}),
 upon exchanging $\theta_1$ and $\theta_2$.

Notice that also in this case, the harmonic function
$h^2$ can be thought of as arising from a profile, but
now the boundary of the droplets are points $\de \D=\{{\bf x_{0,i}}\}$.
The profile function reads:
$$
{\bf x'}(v_i)={\bf x_{0,i}}\quad\quad 0< v_i < Q_i~.
$$

\subsection{Solutions in GMR form}

\label{GMRform}

Given that our geometries are supersymmetric solutions of minimal
supergravity in six dimensions, there must be a change of
coordinates that cast them in the general form presented in \cite{GMR}.
In the present section we give the map between the solutions in
\ref{resume} and the canonical form of \cite{GMR}. As a bonus,
we will show how the condition (\ref{hvsz}) can be lifted by
relaxing the metric ansatz for the torus of isometries allowing
for an off-diagonal term.

Recall that the full six-dimensional metric can be always written
\cite{GMR} as
\begin{equation}
\label{eqn:lineelt}
 ds^2 = 2 H^{-1} \left( du + \beta_m dx^m \right) \left( dv + \omega_m dx^m +
 \frac{\cal F}{2} \left(du + \beta_m dx^m \right) \right) - H h_{mn}
 dx^m dx^n
\end{equation}
with the functions $H,{\cal F}$, the one-forms $\beta,\omega$, and
the four-metric $h_{mn}$ obeying certain coupled differential
equations. It turns out that the solutions of the previous
section fall into the $u$-independent class of that considered in
\cite{GMR}.
More precisely, out of the two null directions,
we can define a time coordinate $t$ and a coordinate $\alpha$ via
\begin{eqnarray}
u  =  \frac{1}{\sqrt{2}} (t-\alpha)\qquad \qquad v  =
\frac{1}{\sqrt{2}} (t+\alpha)~.
\end{eqnarray}
In addition we take:
\begin{eqnarray}
{\cal F} & = & 0\quad \qquad H  =  h^2\\
\beta & = & \frac{1}{\sqrt{2}}(C -zd\phi)\qquad \quad \omega  =
\frac{1}{\sqrt{2}}(C+zd\phi)~.
\end{eqnarray}
The four-metric $h_{mn}$ has to be hyper--K\"ahler \cite{GMR}, and
we take this to be flat, of the form
\begin{eqnarray}
h_{mn}dx^m dx^n & = & dx^2_1 + dx^2_2 + dy^2 + y^2 d\phi^2~.
\label{flatmetric}
\end{eqnarray}
This is rather natural, given that we are interested in geometries
dual to D1D5 systems (without momentum) moving in a flat transverse space.
More general hyper-K\"ahler spaces are relevant for
D1 states on a curved manifold (K3) or D1D5 systems with momentum
\cite{lunin,giusto}.
Note that the definition of $\phi$, has been chosen such that it
has periodicity $2\pi$.
% This can be checked from the Jacobian of
%change of coordinates, which is $2$.
In the new coordinates the metric can be written in the following form
\begin{eqnarray}
ds^2 &=& h^{-2} \left[ (dt+C)^2-(d\alpha+zd\phi)^2\right] - h^2
(dx_1^2+dx_2^2+dy^2+y^2 d\phi^2)\label{inspiring}\\
&=& h^{-2}(dt+C)^2
- h^2 (dy^2+dx_1^2+dx_2^2) - \left[ h^2 y^2+h^{-2}(z+\frac12)^2\right]\,d\theta_1^2\nonumber\\
& -& \left[h^2 y^2+h^{-2}(z-\frac12)^2\right]\,d\theta_1^2
-2\,\left[h^{-2}(\frac14-z^2)-h^2 y^2\right]\,d\theta_1\,
d\theta_2 \label{met12}
\end{eqnarray}
 where in the second line we performed the change of variables
\begin{eqnarray}
\theta_1  =  \alpha + \frac{1}{2}\phi\qquad \qquad \theta_2  =
\alpha - \frac{1}{2}\phi ~.\label{t1t2}
\end{eqnarray}
%leading to:
%\begin{eqnarray}
%ds^2 &&= h^{-2}(dt+C)^2
%- h^2 (dy^2+dx_1^2+dx_2^2) - (h^2 y^2+h^{-2}(z+\frac12)^2)d\theta_1^2\nonumber\\
%&& - (h^2 y^2+h^{-2}(z-\frac12)^2)d\theta_1^2
%-2\,(h^{-2}(\frac14-z^2)-h^2 y^2)\,d\theta_1\,
%d\theta_2~.\label{met12}
%\end{eqnarray}
Note that the $g_{\theta_1\theta_2}$ term in (\ref{met12}) precisely cancels
using the fact that $z$ and $h^2$ are related via (\ref{hvsz}), and the remaining terms recombine
to reconstruct the metric (\ref{themetric}).

It is easy to check that the equations of \cite{GMR} are
satisfied. Note in fact that
\begin{eqnarray}
d\beta & = &\frac{1}{\sqrt{2}}  (dC - dz \wedge d\phi)\\
d\omega & = &\frac{1}{\sqrt{2}}  (dC + dz\wedge d\phi)
\end{eqnarray}
and that
\begin{eqnarray}
(d\beta)^- = (d\omega)^+ =0\label{dbo}
\end{eqnarray}
is equivalent to equations (5.27), (5.30) of that paper. In
particular these equations imply
$$
dC = \frac{1}{y} *_3 dz \qquad \Rightarrow \qquad \Delta_6 \,
\left(\frac{z}{y^2}\right) = 0~.
$$

Here $(~)^\pm$ indicates (anti)-self-dual part with respect to the four
dimensional metric (\ref{flatmetric}). The condition $(d\omega)^+=0$ of
course implies that ${\cal G}^+=0$ as required by ${\cal F}=0$.
The Bianchi identity and Einstein equation reduce correctly to
\begin{eqnarray}
\Delta_4 h^2 & = & 0~.
\end{eqnarray}
This constitutes a check on our solutions, as well as a proof that
they indeed satisfy the Einstein equations. Moreover, this
explains the reason why $h^2$ was previously found to be harmonic
in \emph{four} dimensions.

Finally, it can be checked that the expression for the flux in \cite{GMR} (in the
$u$-independent class):
\begin{eqnarray}
\label{eqn:Gexpand}
G = {1 \over 2} *_4 d h^2
-e^+ \wedge  {1 \over 2} d \omega
+ {1 \over 2} h^{-2} e^- \wedge d \beta - {1 \over 2} e^+ \wedge e^-
\wedge h^{-2} d h^2 
\end{eqnarray}
agrees precisely with the expression for the flux in (\ref{theflux}).
For this, it  is useful to note
\begin{eqnarray}
e^- & = & \frac{h^2}{\sqrt{2}}\left[ h^{-2} (dt+C) + e^{H+G}d\theta_1 + e^{H-G}d\theta_2
 \right]\\
e^+ & = & \frac{1}{\sqrt{2}}\left[ h^{-2} (dt+C) - e^{H+G}d\theta_1 -
 e^{H-G}d\theta_2 \right]~.
\end{eqnarray}

To summarize, we have found that our solutions comprise a
restricted sub-class of the $u$-independent solutions of \cite{GMR}.
This analysis is useful to understand how one can generalize the starting
ansatz, in order to obtain more interesting solutions.

\subsection{Minimal bubbling}
\label{minbub}

The point to notice is that the metric (\ref{met12}) and
3-flux (\ref{eqn:Gexpand}) are
$\ft12$-BPS solution of minimal supergravity for any choice of the
functions $h^2$ and $z$ satisfying \bea
&& dC = \frac{1}{y} *_3 dz \nn\\
&&\Delta_4 h^2  = 0\nn\\
&&\Delta_6 \, \left(\frac{z}{y^2}\right) = 0~,
\label{condi}
\eea
however, crucially, one can now relax the requirement that $z$ and $h^2$ be related
as in (\ref{hvsz}).
This suggests how to recover linearity: we lift the non-linear
relation (\ref{hvsz}) and consider $h^2$ as an independent
function with respect to $z$.
According to (\ref{met12}), this results into a non-trivial
$g_{\theta_1\theta_2}$ component in the metric. Note that in
the case studied in \cite{LLM},
requiring the solution to posses
an $SO(4)\times SO(4)$ isometry, uniquely fixed the factorized form
 of the metric in the internal six dimensional space.
In the case at hand, the presence of an abelian $SO(2)\times SO(2)$ isometry
allows us to retain the same isometry for a generic tilted
$T^2$-torus\footnote{One could consider
the case in which this torus is non-trivially fibered over the external
four-dimensional space, and allow more general terms like $g_{x^i\theta_j}$,
but this goes beyond the scope of this note.}.

 Motivated by the way of writing $C_i$ and $h^2$ for the basic figures
 as the integrals (\ref{zc}) and (\ref{h2v}), it is tempting to define 
 ``bubbling''
of $AdS_3$  by extending these integrals to the boundary of a generic
droplet distribution in the $y=0$ plane.
 This intuition will be confirmed in the next
section where we will show that solutions derived in this way agree with
those found in \cite{LMS,LMM} for D1D5 classical geometries.
These are clearly supersymmetric solutions, and as shown in 
\cite{LMM}, they are also non-singular.

%We thus define a ``minimal bubbling'' as superposition of 
%solutions of minimal
%supergravity, with basic building blocks the circular droplets $(AdS_3)$ and
%the half planes (pp-wave) and $z=\pm \ft12$ over the plane
%\footnote{The study of solutions with conical singularities is also
%interesting in $D=6$ since they correspond to supergravity 
%duals of chiral primaries in the twisted sector of the orbifold CFT
%\cite{LMM}. They can be accommodated in the bubbling
%picture by taking a string profile covering  $m$ times the circle.}. 
%Notice that these figures
%satisfy the condition $ |\de_v{\bf x'}(v)|^2=$const. 
%that ensures that we stay in
%minimal supergravity \cite{lunin}\footnote{Here we are using our identification
%of the D1D5 profiles with the boundary of the droplet configuration.}.

\begin{figure}[ht]
\epsfxsize = 4cm
\centerline{\epsfbox{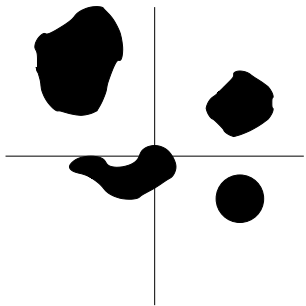}}
\caption{A generic distribution of droplets.}
\label{palleerighe}
\end{figure}

We thus define a bubbling solution of minimal supergravity as specified by 
a droplet distribution ($z=\pm\tfrac{1}{2}$) in the $y=0$ plane.
 In addition we require that the condition $ |\de_v{\bf x'}(v)|^2=$const.
is satisfied. As we will see in the next section, this ensures that
the bubbling is a solution of minimal supergravity, i.e. the dilaton
is constant.   
A generic droplet distribution in the $y=0$ plane is represented in Figure
\ref{palleerighe}. The functions $z$ and $h^2$ are given via the line
integrals (\ref{zc}) and (\ref{h2v}) over the boundary $\partial \D$ of the
filled regions. The harmonic conditions in (\ref{condi}) are then 
automatically satisfied, and the solutions are non-singular.

\begin{figure}[th]
\epsfxsize = 10cm
\centerline{\epsfbox{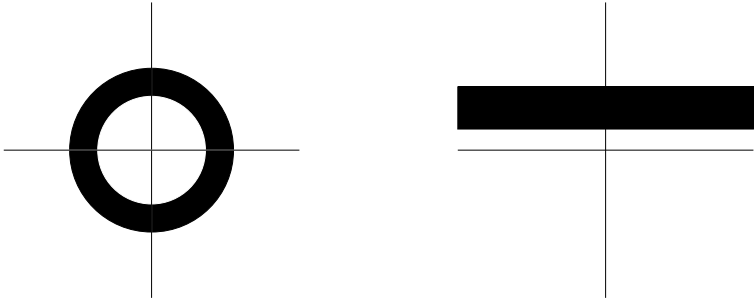}}
\caption{The annulus (or ring) and the strip.}\label{aesfig}
\end{figure}

As an illustration let us consider the annulus diagram\footnote{An equally 
simple example is provided by the strip in Figure \ref{aesfig}.} in Figure 
\ref{aesfig}.

The corresponding profile reads
\bea
{\bf x'}_i(v_i) &=& \left(a_i \cos {v_i\over \xi_i}, a_i \sin {v_i\over \xi_i}\right) 
\qquad  0<v_i<2\,\pi\, \xi_i 
\eea
 with $i=1,2$, and $\xi_1+\xi_2=1$. 
To stay in minimal supergravity we require 
$|\de_v{\bf x'}(v)|^2=a^2$, with $a$ constant, i.e.
$$
{a_1\over \xi_1}={a_2\over \xi_2}=a ~.
$$
 Then the integrals (\ref{zc}), (\ref{zc2}) and (\ref{h2v}) result into:
\bea
z_{\rm ring}({\bf x},y)&=& z({\bf x},y;a_1)-z({\bf x},y;a_2)+\ft12\nn\\
C_{i,{\rm ring}}({\bf x},y)&=& C_{i}({\bf x},y;a_1)-
C_{i}({\bf x},y;a_2)\nn\\
h^2_{\rm ring}({\bf x},y)&=& \xi_1\, h^2({\bf x},y;a_1)+
\xi_2\, h^2({\bf x},y;a_2)\label{annz}
\eea
where the functions $h^2$ and $z$ are those appearing in (\ref{round}), 
and $a_i$ are the radii of the two circles. Notice that by 
construction $z_{\rm ring}$
 is still $\pm \ft12$ over the $y=0$ plane.
 The different signs in $z$ arise from the different orientations of the
 two boundaries. The integral $h^2$ is instead independent of the boundary 
 orientation
 in agreement with the fact that it must be a positive-definite quantity.

The bubbling prescription adopted here is minimal and we don't exclude
other interesting choices. However, as we will see in the next section,
we reproduce all (to our knowledge)
$\ft12$-BPS D1D5 solutions previously known in the literature.
Specifically, we are proposing that \emph {the string profile specifying the
solution in the so-called FP representation, gets identified with the boundary
of a droplet distribution in the dual D1D5 system.}

\section{Adding a tensor multiplet}
\label{addtensor}

As we mentioned in the introduction chiral primaries with
$h=\bar{h}=j=\bar{j}$ appear only in the KK towers descending
from the gravity and tensor multiplets of ${\cal N}=(1,0)$
supergravity. In this section we consider $\ft12$-BPS geometries involving
a non-trivial scalar and anti-self-dual three-form in a tensor multiplet
dressing the minimal supergravity. They are associated to D1D5
geometries (or any of its dual descriptions).

Instead of starting from some ansatz and apply the logic of section
\ref{minimal}
to non-minimal supergravity, we jump at the final result, i.e. we adopt our
bubbling prescription, and check that the results reproduce regular 
solutions.
In fact, as it was realized in section \ref{GMRform},
 the $\ft12$-BPS solutions we are interested in, will lie within the
general class of solutions for a regular distribution of D1D5 branes
\cite{LMS,LMM}.
These solutions are specified by a profile function ${\bf F}(v)$
determining six harmonic functions\footnote{Here we drop ``1''s from the
harmonic functions, as we are interested in near-horizon geometries. Note that 
for pure multi-string solutions, with constant profiles, the ``1'' should be
restored.}
 $f_1$, $f_5$ and $C_i$ in
$\R^4$ \cite{LMM}
\begin{eqnarray}
f_5 =\frac{1}{2\pi}\int_{\partial \D} {dv\over
 |{\bf x}-{\bf x'}|^2+y^2},
~ f_1 =\frac{1}{2\pi}\int_{\partial \D} {|\partial_v{\bf x'}|^2\, dv\over
 |{\bf x}-{\bf x'}|^2+y^2}, \quad
C_i = \frac{\epsilon_{ij}}{2\pi}\,\int_{\partial \D} {\partial_v x'_i(v)\, dv\over
 |{\bf x}-{\bf x'}|^2+y^2}~.\label{profiles}
\end{eqnarray}
 The metric, three-form flux and scalar profiles are given in
 terms of $f_1$, $f_5$ and $C_i$ via\footnote{Notice that the expression for the flux
 (\ref{Gtot}) corrects a minus sign in the $A\wedge B $ term of the flux given 
 in (2.1) of \cite{LMM}. We thank O. Lunin for clarifying this point. 
 Then the map is simply $A_{them}=-\,C_{us}$, and an
 orientation reversal on the 4d base,  $*_{4\,them}=-\,*_{4\,us}$. }
\begin{eqnarray}
ds^2 &=& h^{-2} \left[ (dt+C)^2-(d\alpha+B)^2\right] - h^2
(dx_1^2+dx_2^2+dy^2+y^2 d\phi^2)\label{met}\\
G & = & d \left[ f_1^{-1}(dt+C)\wedge (d\alpha +B)\right] + *_4 d f_5\label{Gtot}\\
dB & = & - *_4 dC \label{dBdC}\\
e^{2\Phi} & = & f_1 f_5^{-1} \qquad\quad h^2=(f_1 f_5)^{1\over 2}~.
\end{eqnarray}
Recall that we are interested in solutions with $J_{12}=j+\bar{j}=2j$,
 $J_{34}=j-\bar{j}=0$, which correspond to having
 an additional $U(1)$ isometry -- the Killing vector being $\de /\de \phi$.
We have used this fact to write the $\phi$-independent harmonic functions
(\ref{profiles}) in terms of the profile ${\bf F}(v)= ({\bf x}'(v),0,0)$.
It then follows from (\ref{dBdC}) that $dB$ is
proportional  to $d\phi$ and hence we can always write
\begin{equation}
B=z d\phi~.
\end{equation}
Now, inserting this into (\ref{dBdC}) we reproduce the equation
(\ref{dCequation}), namely
\begin{eqnarray}
dC & =& \frac{1}{y} *_3 dz
\end{eqnarray}
as well as (\ref{harm6}). After the change of variables
(\ref{t1t2}) the metric can be cast in the usual form
(\ref{met12}) with a (in general) tilted torus fibration.

The supersymmetric solutions are now specified by two
harmonic functions in $d=4$ and one in $d=6$ ,namely
 \bea
&&\Delta_6 \left({z\over y^2} \right)=  0\label{harmz2}\\
%&& \Delta_4 \left( e^{\pm\Phi}\, h^2\right)  =  0
&& \Delta_4  f_i = 0 \qquad \quad i=1,5  ~.\label{harmfi}
 \eea
Note that in \cite{LMM} the authors show that if the harmonic
functions $f_1,f_5,C_i$ are chosen as in (\ref{profiles}),
with a generic profile, the solution is {\em non-singular}. 
This can be used to turn the logic around, and show, using (\ref{harmz2}) 
and the left hand side of (\ref{zc}),  that for all non-singular
profiles specified by (\ref{profiles}) the function $z$ must indeed 
be patch-wise 
$z=\pm \ft12$ on the $y=0$ plane. It would be interesting to derive this
directly from an analysis of the metric singularities like in \cite{LLM},  
and in particular to check whether here more general values of $z$ are allowed.

One can also check that the self-dual three-form $e^{\Phi}G^+$ can
be written as in (\ref{eqn:Gexpand}) and the anti-self-dual
three-form is given by 
\begin{eqnarray}
e^{\Phi}G^- & = & h^2 *_4 d\Phi + e^+\wedge e^- \wedge  d\Phi
\label{genfluxmeno}
\end{eqnarray}
in agreement with the result of \cite{Cariglia:2004kk}.
Note that $G=G^+ + G^-$ is of the form
\bea
G = F_{1} \wedge d \theta_1 + F_2 \wedge d \theta_2~.
\label{genflux}
\eea

Notice that in the extended supergravity, parameterizations
with non-constant velocity $|\partial_v {\bf x'}|^2\neq $const. are allowed.
More precisely, the boundary of the domain $\partial \D$ does not specify 
the solution completely, but one must also specify the velocity along this.

\begin{figure}[th]
\epsfxsize = 8cm
\centerline{\epsfbox{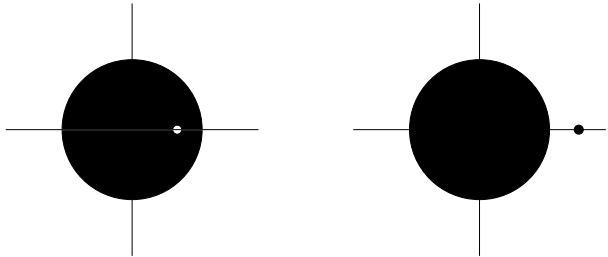}}
\caption{The giant gravitons in  \cite{LMS}.}
\label{thegiants}
\end{figure}

As a simple illustration, let us consider giant gravitons in $d=6$ dimensions
\cite{LMM}.
 The corresponding droplet configurations are depicted in Figure 
 \ref{thegiants}. They arise from superposing a filled
 circle (AdS) and a point (a string). The corresponding profile
of the constituent solutions are  \cite{LMS}
\bea
{\bf x'}(v)&=&(a \cos {v\over \xi},a \sin {v\over \xi}) \quad\quad    0<v<2\pi
\xi\nn\\
{\bf x'}(v)&=& {\bf b}\quad \quad 0<v< 2\pi (1-\xi)
\eea
with ${\bf b}$ a constant vector describing the position of the point. The point
represents a giant graviton extending in  $AdS_3$ or $S^3$ depending on whether
$|{\bf b}|$ is larger or smaller than $a$ \cite{LMM} -- see Figure \ref{thegiants}.

Let us conclude with a final comment. One could wonder whether in the non-minimal
supergravity there exist interesting ``bubbling'' solutions in the case of a
rectangular torus. Although this is not expected, we have explicitly checked
that this does not happen. All supersymmetric solutions
of six dimensional supergravity with a tensor multiplet were
analysed\footnote{Note that the authors of \cite{Cariglia:2004kk} considered
the more general case of gauged supergravities. However, for our purposes
the gauge fields are set to zero.} in \cite{Cariglia:2004kk}, thus extending the
results of \cite{GMR}. Using the results of \cite{Cariglia:2004kk} it is
straightforward to repeat the analysis of section \ref{minimal} in the case
of non-zero scalar field $\Phi$ and an unconstrained flux of the generic form
(\ref{genflux}).
The analysis of the supersymmetry conditions then goes through, essentially because
the self-dual and anti-self-dual parts of $G$ do not mix \cite{Cariglia:2004kk}.
In particular it follows that the metric is exactly as in (\ref{themetric}).
Moreover, equations (\ref{defh2}), (\ref{dCequation}), (\ref{defz}),
still hold true, so that $z$ obeys again (\ref{harm6}). The only difference arises
from the flux and its Bianchi identity reproduces the
harmonic equations (\ref{harmfi}).
However, $h^2$ is again related to $z$ by (\ref{hvsz}), demonstrating that simply
adding a tensors multiplet, but retaining a rectangular torus, \emph{does not}
restore linearity of the equations.

\section{Concluding remarks}
\label{conclusions}

In this note we investigated $\ft12$-BPS classical geometries
arising from bubblings of $AdS_3\times S^3$ and its pp-wave.
This is the supergravity dual of the CFT arising
in the IR on the D1D5 system, and $\ft12$-BPS deformations of $AdS_3\times S^3$
correspond to chiral primaries in the CFT. We have shown that
if one naively mimics the logic of LLM, the resulting set of solutions
share most of the properties of their higher dimensional analogs.
In particular, rather surprisingly,
they are again uniquely specified by a function $z$ which
takes values $\pm\frac{1}{2}$ on a two-dimensional plane.
However, linearity is lost, due to the fact that the Bianchi identity is not
any more automatically satisfied. This leaves room for few possibilities:
$AdS_3\times S^3$, pp-wave and the multi-string solution.

By mapping these solutions to the general form of \cite{GMR}, it becomes
clear how one can generalize the ansatz, in order to restore linearity.
This is accomplished by allowing for a generically tilted torus with
$SO(2)\times SO(2)$ isometries.
The  upshot of this is that $\ft12$-BPS
solutions dual to chiral primaries operators with $h=\bar{h}=j=\bar{j}$
are now specified by $z$ and an additional function $h^2$ harmonic in $\R^4$.
The functions $z$ and $h^2$ are written in terms of integrals over the
 one-dimensional boundary of a droplet distribution  with $z=\pm \ft12$
 in the $y=0$ plane, i.e. harmonics generated by lines of charges.
The geometries  display an interesting fibration
of the symmetry torus over a four-dimensional base, with a special two-dimensional
plane where one-cycles
shrink to zero size, while  keeping the whole geometry regular.

  After adding a tensor multiplet a wide variety of
 previously known D1D5 classical geometries are reproduced. For instance, 
 giant gravitons are reinterpreted as superposition of points over AdS
 disks. Resuming, bubbling works much as in the ten-dimensional case, but
droplet boundaries reveal to be more fundamental than the droplets themselves.

Perhaps this is not very surprising. In fact, the supergravity duals of chiral
primaries of the D1D5 system were constructed in \cite{LMS,LMM}, in terms of a
profile function ${\bf x'}(v)$ arising via a chain of dualities from fundamental
string-momentum solutions \cite{ss1,ss2,cnm}.
Here we identified the profile with the boundary of the
droplets. The results are consistent with a free fermion description of 
the chiral primaries in the two-dimensional CFT \cite{Lunin:2002fw} and it would
be nice to make this correspondence precise.
  Notice that our analysis showed that the Killing spinors are charged
  under the two $U(1)\times U(1)$
isometry and therefore fermions can get masses even if the internal space
is flat.

% %%%%%%%%%%%%%%%% UNCOMMENT THIS PART FOR BIBTEX %%%%%%%%%%%%%%%%
% \bibliography{flussi}
% \bibliographystyle{JHEPmod}
% \end{document}
% %%%%%%%%%%% CUT OUT THE REST IF YOU USE BIBTEX

\vskip 1cm

\subsection*{Acknowledgments}
\noindent We thank Luis Alvarez--Gaume for useful comments and encouragement.
We are grateful to Rodolfo Russo and in particular Gianguido Dall'Agata,
for useful discussions and collaboration at various stages of this work.
We also thank Oleg Lunin for e-mail exchange, and Juan Maldacena for useful
comments. 

\appendix

\section{Derivation of the solutions for a rectangular torus}

\label{appe}

In this appendix we present the detailed derivation of the solution summarized in
\ref{resume}.  We start with the following ansatz for the six-dimensional metric
and the three-form:
\bea
ds^2 &=& g_{\mu\nu}\, dx^\mu dx^\nu -e^{2A} \, d\theta_1^2-
e^{2D} \, d\theta_2^2\nn\\
G &=&F \wedge d\theta_1+\tilde{F} \wedge d\theta_2~.
\eea
We utilize the standard technique of analysing the supersymmetry
conditions encoded in a set of form bi-linears \cite{GMPW}. 
The tensors we consider are
the timelike vector $K$ and the forms defined in (\ref{eq:Xdeco}), which are
related to each other (using self--duality of $X^{i}$) as:
\begin{equation}
\begin{array}{lcl}
\tilde Y^{i} = -e^{-A+D} *_{4} Y^{i},& \quad & Y^{i} = e^{A-D} *_{4}
\tilde Y^{i}  \\[2mm]
\tilde L^{i}=e^{-A-D} *_4 L^i, & \quad &  L^i = -e^{A+D} *_4 \tilde L^i.
\end{array}
\label{eq:Xdual}
\end{equation}
Using the algebraic relations (\ref{eq:null}) and equation 
(2.12) of \cite{GMR}, we obtain the following relations:
\begin{eqnarray}
K \cdot K & = & e^{2A} + e^{2D}\\
L^i \cdot L^j & = & - \delta^{ij}e^{A+D} K\cdot K,\label{eq:normL}\\
K \cdot L^i & = & 0 \\
L^{i\mu}Y^{j}_{\mu\nu} &=& -\epsilon^{ijk} e^{2A} L^{k}_{\nu} -
\delta^{ij}K_{\nu} e^{2(A+D)}\,.
\label{eq:LY}
\end{eqnarray}
It follows that we can use $(K,L^i)$, appropriately normalized, as a privileged
orthonormal frame in four dimensions.
Moreover, using (\ref{eq:normL}) with (\ref{eq:LY}) one can
get explicit expressions for $Y^{i}$ and $\tilde{Y}^i$, in terms of $K$ and $L^i$:
\begin{eqnarray}
Y^{i} &=& -\frac12 e^{-2D} \left(K\cdot K\right)^{-1} \, \epsilon^{ijk}
L^{j}\wedge L^{k} - \left(K\cdot K\right)^{-1} K \wedge L^{i}\\
\label{eq:Yexpr}
%\end{equation}
%An analogous expression can be derived for $\tilde Y^i$:
%\begin{equation}
\tilde Y^{i} &=& -\frac12 e^{-2A} \left(K\cdot K\right)^{-1} \, \epsilon^{ijk}
L^{j}\wedge L^{k} + \left(K\cdot K\right)^{-1} K \wedge L^{i}~.
\label{eq:Ytildeexpr}
\end{eqnarray}

\subsection*{Analysis of the supersymmetry conditions}

We now turn to the differential conditions. The antisymmetric part
of (\ref{vsusy}) gives
\begin{eqnarray}
d K & = & 2(F + \tilde{F})\label{Keq}\\
d e^{2A} & = & - 2 i_K F\label{Aeq}\\
d e^{2D} & = & - 2 i_K\tilde{F}~.\label{Deq}
\end{eqnarray}
The differential conditions on $X^i$ read
\begin{equation}
\begin{array}{rcl}
d \tilde L^i &=& 0\\[2mm]
dL^i &=& \partial_{\theta_2} Y^i - \partial_{\theta_1} \tilde
Y^i\,\\[2mm]
dY^i &=& \partial_{\theta_1} \tilde L^i\\[2mm]
d\tilde Y^i &=& \partial_{\theta_2} \tilde L^i~.
\end{array}
\label{dXi=0}
\end{equation}
Notice that the forms generically depend on $\theta_1, \theta_2$, reflecting the
fact that the Killing spinors are ``charged'' under the corresponding $U(1)$
isometries. Indeed, it can be shown that if one assumes that the forms do not depend
on the angular variables on $S^1\times S^1$, the system does not have non-trivial
solutions.

We now solve this set of equations and find the general
supersymmetric background preserving (\ref{fluxansatz}).
After some algebra, the system (\ref{dXi=0})
is shown to be equivalent to the following set of conditions
\begin{eqnarray}
dL^{i} &=&\frac12 \, e^{-2(A+D)}
\partial_{\theta_{1}}\left(\epsilon^{ijk}L^{j}\wedge L^{k}\right)
\label{dL}\\
d \left(*_3 L^{i}\right) &=& - e^{A+D} h d\left(\frac{1}{h e^{A+D}}\right)
\wedge *_3 L^{i}
\label{dstarL} \\
~ *_3\partial_{\theta_{1}} L^{i}& = & h \, e^{A+D}dL^{i}, \label{stardthe1}\\
~ *_3\partial_{\theta_{2}} L^{i}& = & -h\, e^{A+D} dL^{i}
\label{stardthe2}\\
dC \wedge L^{i} & = & \frac12 d\left(e^{-2A}h^{2}\epsilon^{ijk}L^{j}\wedge L^{k}\right)
\label{dC1}\\
dC \wedge L^{i} & = & -\frac12
d\left(e^{-2D}h^{2}\epsilon^{ijk}L^{j}\wedge L^{k}\right)~.
\label{dC2}
\end{eqnarray}
Compatibility of (\ref{stardthe1}) and (\ref{stardthe2}) shows that
$\partial_{\theta_1} L^i = - \partial_{\theta_2} L^i$.
The equations (\ref{dC1}) and (\ref{dC2}) instead imply that
\begin{equation}
d\left(e^{-2(A+D)}\epsilon^{ijk}L^{j}\wedge L^{k}\right) = 0~.
\label{const}
\end{equation}

A more useful expression which we will use to solve the above
conditions follows from the $(\mu\nu\theta_1\theta_2)$ component of
(\ref{xsusy}):
\begin{eqnarray}
\nabla_\mu L^i_{\nu} & = & \partial_{\mu} (A+D) L^i_{\nu} + F_{\mu}{}^\rho Y^i_{\rho\nu}+
F_{\nu}{}^\rho Y^i_{\rho\nu} + \frac{1}{2} g_{\mu\nu} F_{\rho\lambda}Y^{i\;\;\rho\lambda}.
\label{Leq}
\end{eqnarray}
The antisymmetric part of (\ref{Leq}) gives
\begin{eqnarray}
d (e^{-(A+D)}L^i) & = &  0\,,
\end{eqnarray}
whose general solution is given by
\begin{equation}
L^i = e^{(A+D)}R^{i}_{j}(\theta_1,\theta_2) dx^j.
\label{Li}
\end{equation}
It can be shown that the matrix $R$ must be an $SO(3)$
rotation, and using this, together with the relation
(\ref{eq:normL}) allows us to read off the complete form of
the metric, which we write below:
\begin{eqnarray}
ds^2 = h^{-2}(dt+C)^2 -h^2 \left(dx^2_1 +dx^2_2 +dx^2_3\right) - e^{2A}
\,d\theta_1^2 - e^{2D} \,d\theta_2^2~.
\end{eqnarray}
This now tells us that
\begin{equation}
~ *_3 L^i = \frac12 e^{-(A+D)} h \, \epsilon^{ijk} L^j\wedge L^k,
\label{starL}
\end{equation}
and we can solve the constraints (\ref{dstarL})--(\ref{dL}).
The first condition (\ref{dstarL}) is identically satisfied.
Eq.~(\ref{stardthe1}) determines now the $\theta_i$ dependence of
$L^i$.
After some algebra we get
\begin{equation}
\partial_{\theta_1} R^i_j = R^i_l \epsilon_{jkl}\partial_l e^{A+D}.
\label{dR}
\end{equation}
Since $R$ is an $SO(3)$ matrix, it follows that one of the
three $x^i$ coordinates is
\begin{equation}
x^{3} = y = e^{A+D} 
\label{Hy}
\end{equation}
and we define $e^G=e^{A-D}$.
In this way, $R$ must be a rotation in the other two coordinates by an
angle $\theta_1 - \theta_2$, so that we also solve (\ref{stardthe2}).
%The equation (\ref{dC1}) (or the equivalent (\ref{dC2})) give us an
%expression for $C$:
%\begin{equation}
%dC = *_3 {1\over y} d \left(h^{2}\, y\, e^{-G}\right) = - *_3 {1\over y}  d \left(h^{2}\, y\,
%e^{G}\right).
%\label{dC}
%\end{equation}
%The two are equivalent through the use of (\ref{const}) which now is
%identically satisfied.
(\ref{dL}) is now trivially satisfied.
Next we solve (\ref{Keq}) and (\ref{Aeq})--(\ref{Deq}).
Using the explicit form of $K$ as a form, (\ref{Keq}) reads
\begin{equation}
d (h^{-2}(dt+C)) = 2(F + \tilde{F})~.\label{dKeq}
\end{equation}
Following \cite{LLM}, we pose
\begin{eqnarray}
B &=& B_t (dt +C) +\hat B\\
\tilde B &=& \tilde B_t (dt +C) +\hat {\tilde {B}},
\end{eqnarray}
hence (\ref{Aeq}) and (\ref{Deq}) give
\begin{eqnarray}
 d B_t = \frac{1}{2} d(y \,e^{G})~, \qquad \quad  d \tilde {B}_t=\frac{1}{2}
d( y\, e^{-G}) ~,
\end{eqnarray}
which can be integrated to
\begin{eqnarray}
B_t = \frac{y}{2}\, e^{G}~,\qquad \quad \tilde B_t & = & \frac{y}{2} e^{-G}~,
\end{eqnarray}
and we have set to zero irrelevant integration constants.
Inserting these values into (\ref{dKeq}), one component is identically
satisfied using (\ref{KK}) while the non--trivial part implies
\begin{eqnarray}
d\hat{B} + d \hat{\tilde B} & = & 0~.
\end{eqnarray}
Now consider the 3--form flux. Self-duality implies:
\begin{eqnarray}
d\hat{B}+B_t dC & = &  h^2 e^{G}\, *_3 d\tilde{B}_t\label{self1}\\
d\hat{\tilde{B}}+\tilde{B}_t dC & = & -h^2 e^{-G}\, *_3 d B_t~.\label{self2}
\end{eqnarray}
Summing the two equations (\ref{self1}) and (\ref{self2}) one finds
\begin{equation}
dC~=~2 h^4\, y *_3 dG~=~\frac{1}{y}*_3 \, dz \qquad\quad   z=\frac{1}{2}\tanh G~.
\label{dc}
\end{equation}
Notice that this also solves equations (\ref{dC1}), (\ref{dC2}).
Finally, $d\hat B$ can be read off from either of (\ref{self1})
(\ref{self2}) and reads
\begin{equation}
d\hat{B}=-d\hat{\tilde{B}}=\frac{1}{2}\, y\, *_3 d h^2~.
\end{equation}

\end{document}

%% file: texpref.tex
\newcommand{\cref}[1]{Chapter~\ref{#1}}
\newcommand{\beq}{\begin{equation}}
\newcommand{\eeq}{\end{equation}}
\newcommand{\ba}{\begin{array}}
\newcommand{\ea}{\end{array}}
\newcommand{\bcenter}{\begin{center}}
\newcommand{\ecenter}{\end{center}}

\def\IB{\relax\hbox{$\inbar\kern-.3em{\rm B}$}}
\def\IC{\relax\hbox{$\inbar\kern-.3em{\rm C}$}}
\def\ID{\relax\hbox{$\inbar\kern-.3em{\rm D}$}}
\def\IE{\relax\hbox{$\inbar\kern-.3em{\rm E}$}}
\def\IF{\relax\hbox{$\inbar\kern-.3em{\rm F}$}}
\def\IG{\relax\hbox{$\inbar\kern-.3em{\rm G}$}}
\def\IGa{\relax\hbox{${\rm I}\kern-.18em\Gamma$}}
\def\IH{\relax{\rm I\kern-.18em H}}
\def\IK{\relax{\rm I\kern-.18em K}}
\def\IL{\relax{\rm I\kern-.18em L}}
\def\IP{\relax{\rm I\kern-.18em P}}
\def\IR{\relax{\rm I\kern-.18em R}}
\def\IZ{\relax\ifmmode\mathchoice
{\hbox{\cmss Z\kern-.4em Z}}{\hbox{\cmss Z\kern-.4em Z}}
{\lower.9pt\hbox{\cmsss Z\kern-.4em Z}}
{\lower1.2pt\hbox{\cmsss Z\kern-.4em Z}}\else{\cmss Z\kern-.4em Z}\fi}
\def\II{\relax{\rm I\kern-.18em I}}

\def\sCC{{\kern 0.27em\vrule height1.45ex width0.03em depth0em
          \kern-0.30em\rm C}}
\def\C{{\mathchoice
  {\sCC}
  {\sCC}
  {\kern 0.225em \vrule height1.05ex width0.025em depth0em \kern-0.25em \rm C}
  {\kern 0.180em \vrule height0.78ex width0.02em depth0em \kern-0.2em \rm C}
        }}
\def\sHH{{\rm I\kern-.16em{}H}}
\def\H{{\mathchoice
  {\sHH}
  {\sHH}
  {\rm I\kern-.13em{}H}
  {\rm I\kern-.13em{}H} }}
\def\sNN{{\rm I\kern-.16em{}N}}
\def\N{{\mathchoice
  {\sNN}
  {\sNN}
  {\rm I\kern-.12em{}N}
  {\rm I\kern-.10em{}N} }}
\def\sPP{{\rm I\kern-.16em{}P}}
\def\P{{\mathchoice
  {\sPP}
  {\sPP}
  {\rm I\kern-.12em{}P}
  {\rm I\kern-.10em{}P} }}
\def\sQQ{{\kern 0.27em \vrule height1.45ex width0.03em depth0em
          \kern-0.30em \rm Q}}
\def\Q{{\mathchoice
        {\sQQ}
        {\sQQ}
  {\kern 0.225em \vrule height1.05ex width0.025em depth0em \kern-0.25em \rm Q}
  {\kern 0.180em \vrule height0.78ex width0.020em depth0em \kern-0.20em \rm Q}
        }}
\def\sRR{{\rm I\kern-0.16em{}R}}
\def\R{{\mathchoice
  {\sRR}
  {\sRR}
  {\rm I\kern-0.12em{}R}
  {\rm I\kern-0.10em{}R} }}
\def\sZZ{{\rm Z\kern-0.32em{}Z}}
\def\Z{{\mathchoice
  {\sZZ}
  {\sZZ} 
  {\rm Z\kern-0.3em{}Z}     %.3
  {\rm Z\kern-0.25em{}Z} }}  %.25
\def\ZZZ{{\rm Z\kern-0.24em{}Z}}
\def\sII{{\rm I\kern-0.16em{}I}}
\def\I{{\mathchoice
  {\sII}
  {\sII}
  {\rm I\kern-0.12em{}I}
  {\rm I\kern-0.10em{}I} }}

\def\inbar{\,\vrule height1.5ex width.4pt depth0pt}
\font\cmss=cmss10 \font\cmsss=cmss10 at 7pt

\def\smiley{\hbox{\large$\bigcirc$\hspace{-0.80em}\raise.2ex
\hbox{$\cdot\cdot$}\kern-.61em\lower.2ex\hbox{\scriptsize$\smile$}}\ }
\def\frowny{\hbox{\large$\bigcirc$\hspace{-0.80em}\raise.2ex
\hbox{$\cdot\cdot$}\kern-.635em\lower.2ex\hbox{\scriptsize$\frown$}}\ }

\def\I{{\rlap{1} \hskip 1.6pt \hbox{1}}}

\makeatletter
\let\hangafter\@hangfrom
\makeatother